\documentclass[prl,times,notitlepage,amsmath,amssymb,twocolumn]{revtex4-1}
\usepackage[usenames]{color}
\usepackage{amssymb,hyperref}
\usepackage{grffile}
\usepackage[pdftex]{graphicx}
\usepackage{amsmath, amstext, amssymb, amsfonts, amsxtra}
\usepackage[tbtags]{mathtools}
\usepackage{textcomp}
\usepackage{xspace}
\usepackage{icomma}
\usepackage[short]{datetime}

\def \tr{{\mbox{tr}}}

\def \ell{{d}}

\def \k{{k}}

\def \s{{\sigma}}

\def \avb#1{\,{\langle\,#1\,\rangle}\,\xspace}
\def \av#1{\,{\langle\,#1\,\rangle_\bullet}\,\xspace}

\newcommand{\bra}[1]{\ensuremath{\langle #1 \vert}\xspace}%
\newcommand{\ket}[1]{\ensuremath{\vert #1 \rangle}\xspace}%
\newcommand{\avg}[1]{\ensuremath{\langle #1 \rangle}_\bullet\xspace}%

\newcommand{\plus}{{+}}
\newcommand{\minus}{{-}}

\newcommand{\avdiag}[1]{\ensuremath{\langle #1 \rangle_{\mbox{\small diag.}}}\xspace}%
\newcommand{\avtherm}[1]{\ensuremath{\langle #1 \rangle_{\mbox{\small therm.}}}\xspace}%
\newcommand{\avext}[1]{\ensuremath{\langle #1 \rangle_{\mbox{\small ext.}}}\xspace}%
\newcommand{\shift}[1]{\ensuremath{\,:\!{\hat#1}\!:\,}\xspace}%

\def \figpreamble{(Color online) }

\newcommand{\unige}{D\'epartement de Physique Th\'eorique, Universit\'e de Gen\`eve, 1211
  Gen\`eve, Switzerland}%
\newcommand{\bonn}{HISKP, University of Bonn, Nussallee 14-16, D-53115 Bonn, Germany.}

\begin{document}

\title{Impact of local integrals of motion onto metastable non-equilibrium states}

\author{Peter Barmettler$^1$}%
\author{Corinna Kollath$^{2}$}%
\affiliation{$^1$\unige}%
\affiliation{$^2$\bonn}

\date{\today}
\begin{abstract}
We analyse the stationary behaviour of correlations in a strongly correlated
Bose gas in an optical lattice out of equilibrium.  The dynamics are triggered
by a quench of the interaction starting from the strongly interacting limit
where the system is in a perfect Mott state.  Despite the complete
integrability of our theoretical description, we find seemingly thermal
behaviour for the experimentally measurable correlations at large interactions.
Quite opposed, away from the strongly interacting regime these correlation
functions show highly non-thermal quasi-stationary values.  Both situations are
explained by overlaps of the integrals of motion with the observable and the
initial state in an effective thermal ensemble. The results are obtained using
approximate non-equilibrium Mazur equalities.  The good agreement with the
time-dependent calculations suggest that non-equilibrium Mazur equalities are
an efficient way to calculate short range
correlations for arbitrary integrable models.
\end{abstract}

\maketitle
Recent experimental developments in solid state and atomic physics enable the
preparation of metastable non-thermal states of correlated quantum matter.
Ultrafast optical pulses \cite{Perfetti2006,Cavalieri2007,Basov2011} uncover
aspects of strongly correlated materials which are unreachable by conventional
manipulation such as doping or application of pressure.  In certain systems,
for example in optically excited cuprate superconductors \cite{Fausti2011},
long lived states with remarkable properties can be excited. The
relaxation in materials not only stems from the interaction with the electronic
degrees of freedom, but typically many phononic relaxation channels are
present.  In contrast, non-equilibrium states of ultracold atoms
\cite{Greiner2002,Kinoshita2006,Jordens2008,Chen2011a,Trotzky2011,Gring2011,Cheneau2012},
can evolve coherently over long times with only little dissipation into
external environments. Motivated by the experimental realizations, the theoretical description of such non-equilibrium
phenomena and in particular of metastable non-thermal states has attracted a
lot of interest recently.  Several time-dependent numerical and approximative
methods have been developed in the past decade. However, accessing long-time
properties remains still an open challenge.

A special role in the understanding of non-equilibrium states take integrable
models.  
For our purposes it is suitable to define
an integrable system as quantum model exhibiting a macroscopic number of local
operators which commute with the Hamiltonian.  We classify operators as local
if they act on a finite number of lattice sites. Different classes of such
integrable models are quadratic systems, Gaudin-type models
\cite{Faribault2009,BarmettlerGritsev2013} and Bethe ansatz solvable models
\cite{EsslerBook}. Also the few body problems derived by perturbative
treatments (e.g.~\cite{Kollar2011}) can be seen as exactly solvable systems. It
is important to note that a real physical system will never be integrable in a
strict sense. Exactly solvable models should be seen as a tool to effectively
describe metastable non-equilibrium states.  In principle, the knowledge of the
complete set of integrals of motions would enable the exact treatment of
non-equilibrium states of exactly solvable models.  The standard approach is
the so-called 'generalized Gibbs ensemble' (GGE) \cite{RigolOlshanii2006},
which implements constraints due to conservation laws by means of Lagrange
multipliers. However, the usage of the GGE can be difficult and it has so far
mainly been used for quadratic systems (e.g.
\cite{BarthelSchollwoeck2008,Rossini2009,Calabrese2011,Fagotti2013}). For
non-trivially integrable models the GGE has so far only been evaluated in
special cases \cite{Mossel2012a,Kormos,Iyer2013,Fagotti2013b} or for weakly
perturbed states \cite{Fagotti,Pozsgay2013}.  

In this work we will make use of conserved quantities in order to gain further
insights into the non-equilibrium states of integrable systems and to calculate
their properties efficiently.  The main idea is to approximate expectation
values by a truncated series of projections onto independent integrals of
motions. This can be realized by using an adaptation of Mazur bounds
\cite{Mazur,Suzuki1971} to the non-equilibrium regime \cite{Sirker2013}. In
Ref.~\cite{Sirker2013} a condition for thermalization has been derived using
these Mazur-type equalities. We
specifically investigate a strongly interacting one-dimensional Bose gas in an
optical lattice for which metastable states after an interaction quench have
been predicted theoretically \cite{Kollath2007, Roux2009,Biroli2010} and
observed in experiment \cite{Cheneau2012}.  Within an effective integrable
model, we find that only few conserved quantities are sufficient to describe
relevant physical properties.  Besides its conceptual importance, an advantage
of our strategy is, that it is not restricted to quadratic models. It can be
straightforwardly extended to non-trivially exactly solvable systems, such as
one-dimensional fermions and spin chains.  The long-time expectation value is
expressed in terms of thermal averages of observables and integrals of motions,
which (unlike the time-evolution) can be calculated efficiently by standard
equilibrium methods such as the density matrix renormalization group (DMRG) or
exact diagonalizations.

\begin{figure}[ht!]
  \begin{center}
    \includegraphics[width=0.49\textwidth]{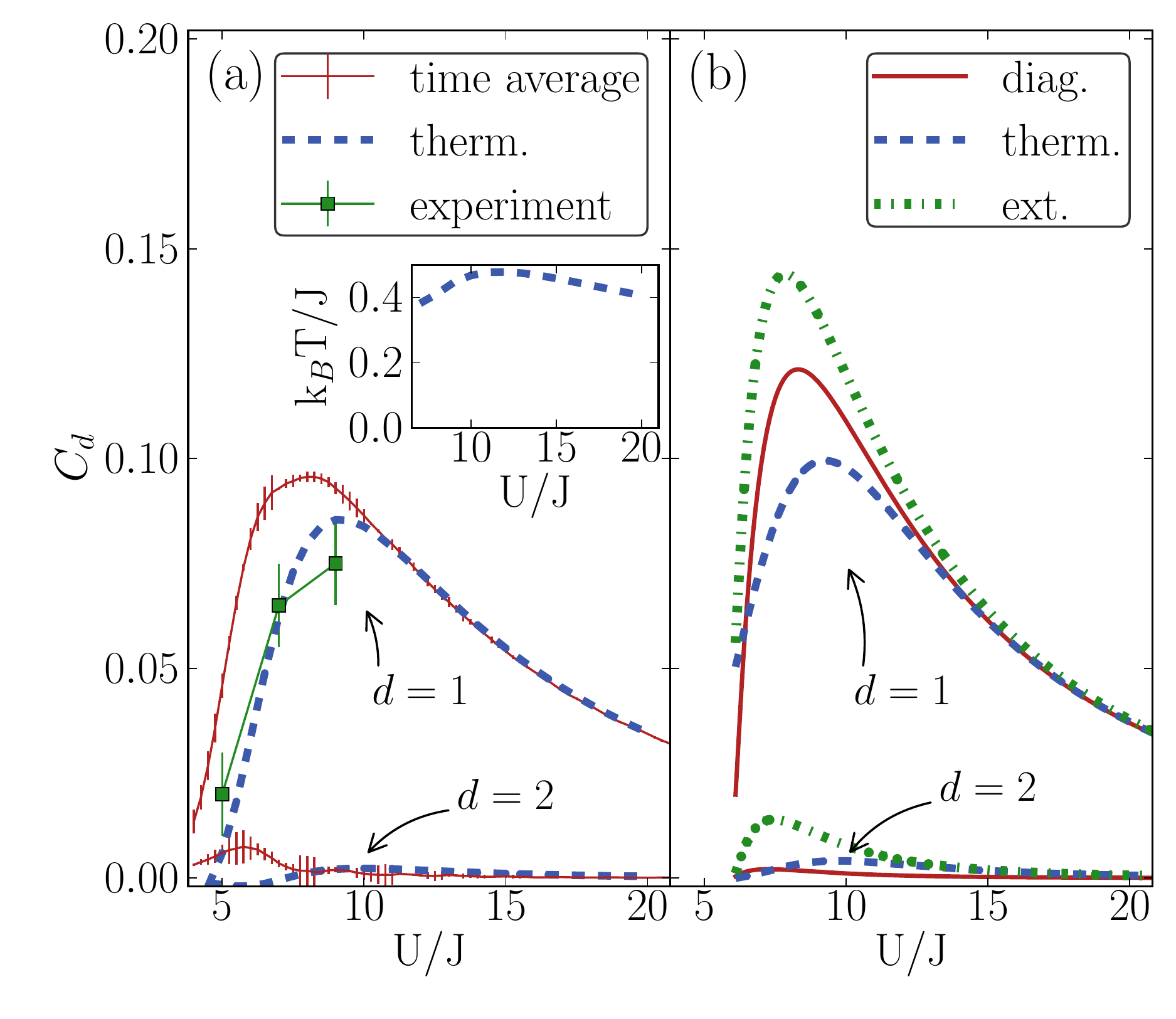}
  \end{center}
  \caption{\figpreamble \label{fig:cdmrgexp} 
  Comparison of different averages for the parity-parity correlations for $d=1$
  (upper curves) and $d=2$ (lower curves). (a) DMRG results obtained by a real-time evolution averaging
  over the time interval $\hbar/J\leq t\leq 3\hbar J$ (error bars reflect the
  mean square deviations from the average) and corresponding thermal expectation
  values. Experimental results are taken from Ref. \cite{Cheneau2012}. The inset shows the effective temperature. (b)
  Results for different ensembles averages (See text) derived from the
  approximate quadratic model \eqref{eq:hfermionic}.}
\end{figure}

Our starting point are bosonic atoms in a 1D tube subjected to a deep
longitudinal optical lattice. This situation is accurately described by the
Bose-Hubbard model $\hat H = \sum_{j} \left\{ -J \,(\hat a^{\dagger}_{j} \,
\hat a_{j+1} + \text{h.\,c.}) + \frac{U}{2}\hat n_j (\hat n_{j}- 1) \right\}$
\cite{JakschZoller1998}. $\hat a^{(\dagger)}_{j}$ is a bosonic annihilation
(creation) operator on site $j$, $\hat n_j=\hat a^{\dagger}_{j}\hat a_{j}$
counts the number of bosons. Interaction $U$ and tunneling $J$ can be
determined from first principles \cite{JakschZoller1998}. The tunneling is
exponentially suppressed upon increasing the intensity of the longitudinal
lattice beam.  Therefore, using a large lattice depth an atomic Mott state with a
single localized atom per site $\ket{\psi_0}=\prod_j \hat
a^{\dagger}_{j}\ket{\mbox{0}}$ can be prepared with high fidelity
\cite{Endres2011,Bakr}. Here $\ket{\mbox{0}}$ is the vaccum state. We work at
filling one and assume a homogeneous system in the thermodynamic limit. For the
times considered here effects of trapping potentials are weak
\cite{Cheneau2012}.

Paired doublons and holes start propagating with well defined velocities in
opposite directions when suddenly ramping down $U/J$ from an effectively infinite value
to a finite one. This gives  rise
to experimentally detectable parity correlations $C_{j-j'}=\avb{\hat p_j \hat
p_{j'}}-\avb{\hat p_j}\avb{\hat p_{j'}}$, $\hat p_j=e^{i\pi\left(\hat
n_j-1\right)}$ \cite{Endres2011}.  
These correlations exhibit a pronounced propagation front  at $t\sim
\hbar(j-j')/6J$ \cite{Cheneau2012,Barmettler2012}. After the passage of
this front, they show damped oscillations around a finite value.  At
large $U/J$ it can be shown that the envelope of the oscillations
decays algebraically as $t^{-3/2}$ for the nearest neighbour and
$t^{-3}$ for the longer distance correlations \cite{Barmettler2012}.
With time-dependent DMRG simulations (See also \cite{Barmettler2012}),
we can study the evolution of the parity correlations up to
$t\sim3\hbar/J$. If interactions are sufficiently strong ($U/J> 4$) a
meaningful stationary value can be extracted from the time interval
$1\hbar/J\leq t\leq3\hbar/J$. The result of this time average is
displayed in Fig. \ref{fig:cdmrgexp}. Only short range pair correlations show
significant magnitudes and we focus on
$C_{d=1}$ and $C_{d=2}$.  The time average of $C_{d=1}$ vanishes in the limit
of $U/J\rightarrow\infty$. It increases when lowering $U/J$ before dropping
relatively quickly to zero after $U/J\sim 8$.  $C_{d\geq2}$ behave similarly,
but at a much lower amplitude. From the experiment
\cite{Cheneau2012} the $C_{d=1}$ stationary values can be extracted.  These lie
somewhat below the theoretical predictions.  This discrepancy can be fully
accounted for by defects in the initial state of the experimental realization.

Under ergodic assumptions \cite{Neumann1929}, one expects that the large-time
limit is described by the Boltzmann distribution $\hat \rho=e^{-\beta
\left(\hat H -\mu \sum_j \hat n_j\right)}/Z$, where $\beta$ and $\mu$ are
chosen such that $\bra{\psi_0}\hat H\ket{\psi_0}\stackrel{!}{=}\tr ~\hat H \hat
\rho$ and $\sum_j\bra{\psi_0}\hat n_j\ket{\psi_0}\stackrel{!}{=}\sum_j\tr~ \hat
n_j \hat \rho$. This thermal ensemble is simulated using the finite-temperature
DMRG method \cite{Schollwock2011}. The obtained effective temperature (Fig. 1(a) inset)
is relatively large $k_B T\gtrsim 0.4J$ and varies little in the shown region $5<U/J<20$.
As evident in Fig.~1(a), there are significant differences between thermal and
time-averaged ensembles, but qualitatively they correspond unexpectedly well
(see also \cite{Biroli2010}). In particular the thermal average shows the same basic
features:
At large interactions coherent excitations only emerge
perturbatively on the order $J^2/U$ \cite{Huber2007,Barmettler2012} and
parity correlations become highly suppressed at these high effective
temperatures. The vanishing correlations at weaker interactions are mainly due
to the fact that correlations are no more purely doublon-hole-like, but mix with
doublon-doublon and holon-holon correlations. 
In the strongly interacting regime $U/J\geq12$ the thermal values even
quantitatively reproduce the time-averaged ones. 
In contrast, at intermediate interactions the
non-equilibrium values are significantly larger than the thermal ones. In the
following we analyze this crossover between apparently thermal and non-thermal
regimes and relate it to the sensitivity of observables to conserved
quantities.

The considered stationary state can be described approximately by a quadratic model of fermionized doublons and
holes. Following Ref. \cite{Barmettler2012}, this model can be derived by first truncating the local
Hilbert space to contain maximally two atoms per site, i.e. the site basis is
spanned by the states $\ket{0},\ket{1},\ket{2}$. Further, Jordan-Wigner
fermion operators are introduced to describe the doublons $\hat c_{j,\plus}^\dagger\propto \ket{2}\bra{1}$
and holes $\hat c_{j,\minus}^\dagger\propto \ket{0}\bra{1}$ (the atomic Mott
state at filling one $\ket{\psi_0}$ is a vacuum of auxiliary fermions in this representation).
By retaining only quadratic terms and neglecting a constant energy shift the Bose-Hubbard model reduces to:
\begin{multline} \label{eq:hfermionic} \hat H=\sum_j\bigg\{ -2J \hat
		c_{j,\plus}^\dagger \hat c_{j+1,\plus}\,-\,J\hat
		c^\dagger_{j+1,\minus}\hat c_{j,\minus}\\ -J\sqrt{2} \left(
		\hat c_{j,\plus}^\dagger \hat c^\dagger_{j+1,\minus}\,-\,\hat
		c_{j,\minus}\hat c_{j+1,\plus} \right) +\text{h.c.}\\
		+\frac{U}{2}(\hat n_{j,+}+\hat n_{j,-}) \bigg\}\,.
	\end{multline} 
The Hamiltonian neglects the constraint that multiple
occupancies of different auxiliary species on the same site are not allowed. This leads
to a systematic overestimation of the number fluctuations, but
otherwise reproduces well the physics of the intermediate time regime.
The Hamiltonian \eqref{eq:hfermionic} is diagonalized in momentum space
by using Bogolyubov transformations, $\hat \gamma_{k,\sigma}=u_k \hat
c_{\k,\sigma}- v_k \hat c^\dagger_{-\k,-\sigma}$
with $\sigma=\pm$,
$u_k=\cos\left(\text{atan}\left(\frac{4J\sqrt{2}\sin(\k)\,,}{-6J\cos(\k)+U}\right)/2\right)$
and $v^2_k=u^2_k-1$. In this representation $\hat H=\sum_{\k,
\sigma}\epsilon_\sigma(k)\hat \gamma_{k,\sigma}^\dagger\hat
\gamma_{k,\sigma}$ where the quasiparticle energies are given by
$\epsilon_\sigma(k)=-\sigma
J\cos(\k)+\frac{1}{2}\sqrt{\left(-6J\cos(\k)+U\right)^2+\left(4J\sqrt{2}\sin(\k)\right)^2}$.

Unlike in DMRG simulations, it is possible to access the long-time limit within
the quadratic fermionic model.  We will evaluate the quadratic fermionic model
in different ensembles. For all of them, two-point correlations are solely
determined from momentum distributions for auxiliary fermions: 
\begin{align}
\avg{\hat
c_{k,\sigma}^\dagger \hat c_{k,\sigma}}=u^2_\k
f_{\k,\sigma}-v^2_\k(1\!-\!f_{\k,-\sigma})\,,
\label{eq:equalcorr}\notag\\
\avg{\hat c_{k,\sigma}\hat c_{-k,-\sigma}}=u_\k
v_\k\left(f_{\k,\plus}\!+\!f_{\k,\minus}\!-1\!\right)\,,
\end{align}
where the quasiparticle distributions $f_{\k,\sigma}=\avg{\hat
\gamma^\dagger_{k,\sigma}\hat \gamma_{k,\sigma}}$ are evaluated in
a given ensemble ($\bullet$ is a placeholder for the type of
the ensemble).  For example, in the long-time limit after
the quench from the perfect Mott state expectation values are
determined by the {\it diagonal } ensemble
\cite{BarthelSchollwoeck2008}: 
\begin{align}
	\label{eq:mddiag}
	\avdiag{\hat\gamma^\dagger_{k,\sigma}\hat\gamma_{k,\sigma}}&:=-v^2(k)\,,
\end{align}
which is equivalent to the 'generalized Gibbs
ensemble' \cite{RigolOlshanii2006}. In the {\it thermal}
ensemble we fix the energy and the total number of particles to
the initial values by an effective temperature and a chemical
potential.  This leads to the Fermi-Dirac distributions,
\begin{align}
	\avtherm{\hat\gamma^\dagger_{k,\sigma}\hat\gamma_{k,\sigma}}&:=
	n^F_\beta(\epsilon_{\sigma}(k)-\sigma\mu)\,,
\end{align}
with $n^F_\beta(\epsilon)=1/(e^{\beta
\epsilon}+1)$, which is fundamentally different from the quasiparticle
distribution of the diagonal ensemble \ref{eq:mddiag}.

We also introduce a natural {\it extended}
ensemble which fixes energies and number operators for both
species individually: 
\begin{align} \label{eq:avext}
	\avext{\hat\gamma^\dagger_{k,\sigma}\hat\gamma_{k,\sigma}}&:=n^F_{\beta_\sigma}(\epsilon_{\sigma}(k)-\sigma\mu_\sigma)\,.
\end{align} The effective inverse temperatures $\beta_\sigma$
and chemical potentials $\mu_\sigma$ are chosen such that
expectation values of
$\sum_k\epsilon_{\sigma}(k)\hat\gamma^\dagger_{k,\sigma}\hat\gamma_{k,\sigma}$
and $\sum_k\hat\gamma^\dagger_{k,\sigma}\hat\gamma_{k,\sigma}$
are equal to those of the initial state. 

The parity correlations in the diagonal, thermal, and extended ensembles for the
quadratic approximation \eqref{eq:hfermionic} are shown in Fig 1b.  As demonstrated
previously in time-dependent simulations \cite{Barmettler2012}, the
correlations are somewhat overestimated as compared to exact DMRG simulations
for the original Bose-Hubbard models (displayed in Fig. 1a). Apart from that it
is justified to explain the properties of the metastable state in the
Bose-Hubbard model within the quadratic model \eqref{eq:hfermionic}.

As in the full Bose-Hubbard model (see Fig. 1a), the diagonal ensemble of the
effective integrable model exhibits a crossover between a seemingly thermalized
behavior at large $U\gtrsim15$ and non-thermal expectation values at smaller
interactions. This behaviour is found for both the thermal ensemble and the
extended ensemble.  Whereas the thermal ensemble underestimates the
correlations compared to the diagonal ensemble, the extended ensemble
overestimates them.  We find that this crossover is rather robust and occurs
also for larger-distance correlations and other observables such as single
particle and string \cite{Endres2011} correlations. Other quantities, such as
momentum distributions of quasiparticles, differ from their thermal values.  A similar effect has
been observed in a quantum Ising chain \cite{Rossini2009}, where off-diagonal
correlators approach thermal values while diagonal ones do not. 

In order to understand such behaviour, we propose to use a hierarchy of 
conserved quantities out of which only few turn out to be relevant. In the
context of the generalized Gibbs ensemble one typically resorts to the
occupancies $\hat\gamma^\dagger_{k,\sigma}\hat\gamma_{k,\sigma}$. In the
considered quadratic model this ensemble is equivalent to the diagonal ensemble
and therefore does not provide new physical insight.  A promising idea has been
put forward by Fagotti et al.~\cite{Fagotti2013} who used a {\it truncated GGE}
formed by a limited number of {\it local} integrals of motions
\eqref{eq:integrals}.  This is a valuable approach but the evaluation of this
ensemble is technically difficult especially when non-quadratic models shall be
considered.   Similarly to Ref. \cite{Fagotti2013} we use local integrals of
motions but we avoid the evaluation of the truncated GGE by expanding the
observables in a series of projections onto these local integrals of motions. Such
expansion technique has been developed by Mazur \cite{Mazur} and is employed to
estimate the influence of the conserved quantities to response functions
\cite{Mazur,Suzuki1971,Zotos1997,Zotos1999,Caux2011,Sirker2011}. For
non-equilibrium setups, a Mazur-type equality has been used to estimate finite
size effects \cite{Caux2011} and a condition for thermalization \cite{Sirker2013}.

We construct the local integrals of motion in the effective model using
 \begin{align} \label{eq:integrals}\hat
	I_{\sigma,\alpha}=\sum_{\k}\cos\left(\alpha
	k\right)\hat\gamma_{\k,\sigma}^\dagger\hat\gamma_{\k,\sigma}\,,
\end{align} for $\alpha \geq 0$.  These quantities are linear combinations of integrals of motion and thus remain 
integrals of motions. In addition, it is convenient to define $\hat
I_{\sigma,-1}=\sum_k\epsilon_{\sigma}(k)\hat\gamma^\dagger_{k,\sigma}\hat\gamma_{k,\sigma}$.

The hierarchy of integrals of motions should be such that lower integrals are
not contained in the higher ones. This can be achieve by orthogonalizing the
integrals One can use different ensemble averages as a measure for the overlap
of conserved quantities and we formulate our procedure for a general ensemble
$\av{\cdot}$. In order to achieve a set of orthogonal integrals we first shift
expectation values of integrals of motions to zero,
$\shift{I_{\sigma,\alpha}}=\hat I_{\sigma,\alpha} - \av{\hat
	I_{\sigma,\alpha}}$  and then apply the Grahm-Schmidt scheme
\begin{align}
	\hat J_{\s,-1}&=\shift{I_{\s,-1}}\notag\\
	\hat J_{\s,0}&=\shift{I_{\s,0}}-\av{\shift{I_{\s,0}} J_{\s,-1}}\hat J_{\s,-1}/\av{\hat J_{\s,-1}^2}\notag \\
	\vdots&\,\,\,\,\,\,\,\,\,\,\vdots
\end{align} 
such that 
\begin{align} 
	\av{\hat J_{s,\alpha}\hat J_{s',\alpha'}}&=\delta_{s,s'}\delta_{\alpha,\alpha'}\av{J_{s,\alpha}^2},\mbox{  and}\notag\\
	\av{J_{s,\alpha}}&=0\,.
\end{align} 
With this transformed set, we can closely follow Mazur's arguments
\cite{Mazur,Suzuki1971} and expand the observable in the diagonal
ensemble in the conserved quantities \cite{Sirker2013}: \begin{align}
\avdiag{\shift{O}}=\sum_{\s,\alpha}\av{\hat O\hat
	J_{\s,\alpha}}\tr\left(\hat\rho_0 \hat J_{\s,\alpha}\right)/\av{\hat
		J_{\s,\alpha}^2}\label{eq:mazur}\,.  
\end{align} 

This modified Mazur equation measures the distance between the diagonal and an approximate
ensemble $\av{\cdot}$ for a given quantity $\hat O$ (We recall that
$\avdiag{\shift{O}}=\avdiag{\hat{O}}-\av{\hat{O}}$).  In the original
linear response formula of Mazur \cite{Mazur,Suzuki1971} 
all terms are positive and the
equation turns into an inequality when the series is truncated. In the present
case such an inequality cannot be derived.  However, in practice there is nevertheless no
need to evaluate the entire series. Strictly speaking, Eq.~\eqref{eq:mazur} is
only exact when all local \eqref{eq:integrals} and non-local integrals of
motion (i.e.~projections onto eigenstates) are included. But on rather
general grounds, one can argue that eigenstates become irrelevant for local
observables in the thermodynamic limit
\cite{Caux2011,Sirker2013}. For short range correlations one can even
go one step further and truncate the series to contain few terms. In the
considered case the amplitudes of the contributions of higher order terms decay
rapidly and already the first few are sufficient to obtain accurate results.  
\begin{figure}[t!]
  \begin{center}
    \includegraphics[width=0.4\textwidth]{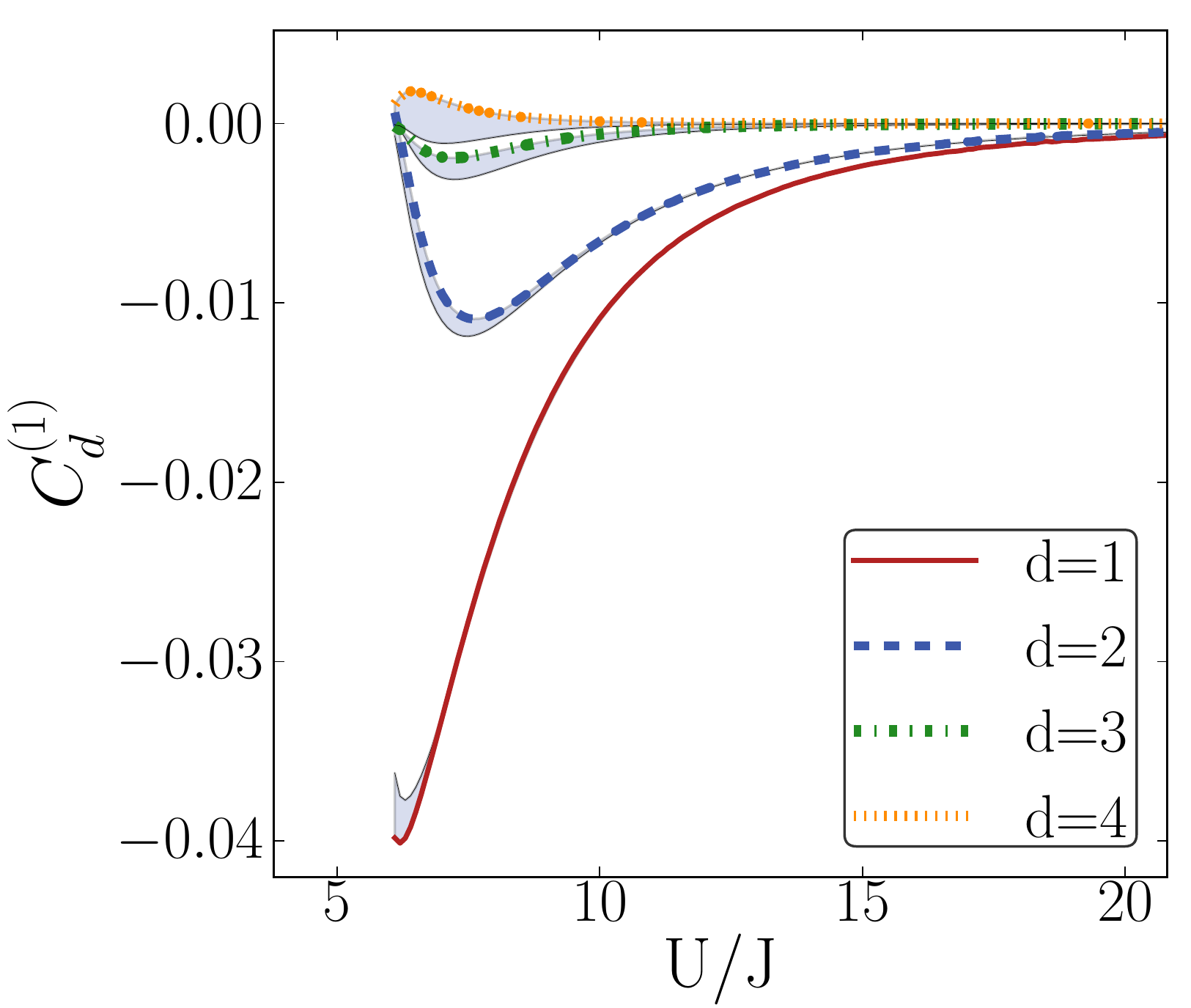}
  \end{center}
  \caption{\label{fig:gpmimo} Contribution from the first non-trivial integrals
  of motion to the correlation function, $C^{(1)}_d$, Eq. \eqref{eq:imo}, are plotted as various
  dotted curves. The offsets of $C^{(1)}_d$ to the differences between extended and
  diagonal ensembles are displayed as shaded regions. 
  }
\end{figure}

To demonstrate this we stick to the extended ensemble
$\av{\cdot}=\avext{\cdot}$ in which all terms with $\alpha\leq 0$ in Eq.
\eqref{eq:mazur} vanish by definition. For the experimentally relevant
parity-parity correlations the first relevant term in the Mazur expansion is
\begin{align}
\label{eq:imo}
C^{(1)}_d:=\sum_{\sigma}\avext{:\!\hat p_j\hat p_{j+d}\!:\,\hat J_{\sigma,1}}\tr\left(\hat\rho_0 \hat J_{\sigma,1}\right)/\avext{\hat J_{\sigma,1}^2}\,.
\end{align}
As shown in Fig. \ref{fig:gpmimo} for small distances $d$, this term almost
fully describes the deviation from the extended thermal ensemble. Only at
distances $d\geq 4$ higher terms become significant. For large distances or
higher precision one could include higher integrals of motion $\alpha\geq1$.

With this result we can interpret the crossover from thermal to non-thermal
behaviour upon decreasing interaction (See Fig. 1) in terms of equilibrium
expectation values of integrals of motions and the observable: Both, overlaps
of the integral of motion with the observable $\avext{:\!\hat p_j\hat
	p_{j+d}\!:\,\hat J_{\sigma,1}}$ and the initial state
	$\tr\left(\hat\rho_0 \hat J_{\sigma,1}\right)$ are negligibly small at
large $U$ and small $d$ and become significant only at weaker interactions. It is
important to note that this effect goes beyond second order perturbation theory -- even in the
strongly interacting regime the $\frac {J^2}{ U}$ expansion
\cite{Huber2007, Barmettler2012} is not sufficient to describe the
observed behaviour. 
The effective integrals of motion $\hat J_{\sigma,1}$ are
fermionic two-point correlators which mainly extend over distances $d=1$ and $d=2$. 

For future investigations, it will be interesting to approach also
non-quadratic integrable systems. In Ref. \cite{Tsuji2013} for example, a
weak interaction quench in the 1D Hubbard model for fermions has been studied. This model is
integrable by the Bethe ansatz \cite{LiebWu1968}.  The time-dependent DMRG
simulations show a relatively quick relaxation \cite{Tsuji2013} of the double
occupancy to a value which is not completely far off the thermal one. The first
non-trivial integral of motion for the 1D fermionic Hubbard model is an energy
current formed by operators on three neighboring sites \cite{Shastry1986},
which is similar to $\hat I_{\sigma,1}$ used here. Therefore, we expect that
mainly the first integral of motion will contribute to the local double
occupancy. Overlaps with the conserved current in the modified Mazur equation \eqref{eq:mazur}
can be accessed by finite temperature DMRG or other methods such as high-temperature
series expansions, allowing to make predictions for the long time limit which
is inaccessible to time-dependent DMRG simulations.  Another interesting
perspective is the calculation of the effect of non-integrable perturbations
within the Mazur expansion. This may be a feasible way to extract
thermalization time scales in near-integrable systems. 

\paragraph{Acknowledgements -- } We would like to acknowledge discussions with M. Cheneau, L. Foini and D. Fioretto and thank M. Cheneau for providing the experimental data. Computer simulation employed the alps libraries \cite{Bauer2011}.

\bibliographystyle{prsty}

\bibliography{refs}
\end{document}